\documentclass{webofc}

\usepackage[varg]{txfonts}   
\usepackage{hyperref}
\usepackage{url}
\usepackage{lineno}
\hypersetup{colorlinks=true,citecolor=blue,urlcolor=blue,linkcolor=blue}
\begin{document}
%
\title{Measurement of charge-dependent directed flow in STAR Beam Energy Scan (BES-II) Au+Au and U+U collisions}
%
%

\author{\firstname{Muhammad Farhan} \lastname{Taseer}\inst{1,2}\fnsep\thanks{\email{mfarhan_taseer@impcas.ac.cn}} for the STAR Collaboration
}

\institute{Institute of Modern Physics, Chinese Academy of Sciences, Lanzhou 730000, China 
\and 
University of Chinese Academy of Sciences, Beijing 100049, China}

\abstract{We presented the rapidity dependence of directed flow ($v_1$) and its slope ($dv_1/dy$) for $\pi^\pm$, $K^\pm$ and $p(\bar{p}$) in Au+Au collisions at  $\sqrt{s_{NN}}$ = 7.7 -- 19.6 GeV from the Beam Energy Scan Phase-II, as well as in U+U collisions at $\sqrt{s_{NN}}$ = 193 GeV measured by the STAR experiment. The $v_1$ values are reported as a function of rapidity and centrality. Additionally, the $dv_1/dy$ and the charge dependent difference, $\Delta(dv_1/dy)$, of the three identified particles in U+U collisions is compared to those in Au+Au and isobar (Ru+Ru and Zr+Zr) collisions. These findings offer insights into the initial electromagnetic field as well as baryon transport at various system sizes and beam energies.}

\maketitle
\section{Introduction}
\label{intro}
An ultra-strong magnetic field ($B\approx10^{18} \text{ Gauss}$) is anticipated during the early stages of heavy-ion collisions. Such a strong magnetic field holds significant importance in exploring the properties of QCD matter under extreme conditions, including understanding topology of QCD vacuum and QCD phase transition. Indeed, magnetic fields are primarily generated by spectators and dissipate rapidly, with timescales similar to the duration of the colliding nuclei passage \cite{1, 2}. The presence of such an extremely strong magnetic field significantly influences the properties of the particles produced in the initial stage.  
\medbreak
The directed flow or the first harmonic flow coefficient ($v_1$) describes the collective sideward motion of produced particles and nuclear fragments, and carries information from the very early stages of collision \cite{3}. Specifically, particles and antiparticles with opposite charges experience different contribution of the electromagnetic forces to their rapidity-odd directed flow, $v_1(y)$ \cite{4}. In addition, $v_1$ can capture information from the initial geometry of the system and also offer means to understand baryon transport in the early phase of collision.  
\medbreak
In the expanding quark-gluon plasma (QGP), quarks experience different electromagnetic effects. Specifically, the Lorentz force \textbf{F} = q\textbf{v} × \textbf{B} (resulting from the Hall effect) acts in the opposite direction to the forces arising from Faraday induction and Coulomb effect \cite{1, 2}. The $\Delta(dv_1/dy)$, difference in $v_1$ slope between positively and negatively charged particles, can be used to investigate the effects of electromagnetic fields in heavy-ion collisions. Moreover, model studies \cite{5,6,7} suggest that transported quarks from the colliding nuclei to the mid-rapidity regions and the produced quarks in the final state may have different contributions due to differences in quark compositions, resulting a charge dependent $v_1$ and $\Delta(dv_1/dy)$ \cite{4}.
\medbreak
Previous studies by STAR Collaboration on $\Delta(dv_1/dy)$ in peripheral Au+Au collisions at $\sqrt{s_{NN}}$ = 27 GeV, as well as Au+Au and isobar collisions at top RHIC energy ($\sqrt{s_{NN}}$ = 200 GeV) are consistent with the expectation from the dominance of Faraday and Coulomb effects in peripheral collisions \cite{4,8}. Another STAR analysis \cite{9} observed the positive $\Delta(dv_1/dy)$ splitting in the mid-central collisions using combinations of hadrons without transported quarks, which can be explained by the presence of an electromagnetic field predominantly influenced by the Hall effect. We extend these charge dependent $v_1$ measurements by different collision systems such as U+U, Au+Au and isobar (Ru+Ru and Zr+Zr).

\section{Dataset and event selection}
The data set analyzed in this work was recorded by the STAR experiment at the Relativistic Heavy Ion Collider (RHIC) in the year 2012 for U+U collisions at $\sqrt{s_{NN}}$ = 193 GeV. A total of 250 million good quality minimum-bias triggered events were analyzed, with each event required to have a primary vertex within |$V_{z}$| $\le$ 50 cm along the beam direction and |$V_{r}$| $\le$ 2 cm in the transverse direction. Time Projection Chamber (TPC) of STAR is used for charged particle tracking within pseudorapidity |$\eta$| < 1 and provides full 2$\pi$ azimuthal coverage. Standard track selection and particle identification, utilizing TPC and Time-of-Flight (TOF) detectors, are performed as described in Ref. \cite{4}.
\section{Results and discussion}
\vspace{-8pt} 
\begin{figure}[h]
\centering
\includegraphics[width=9.5 cm,clip]{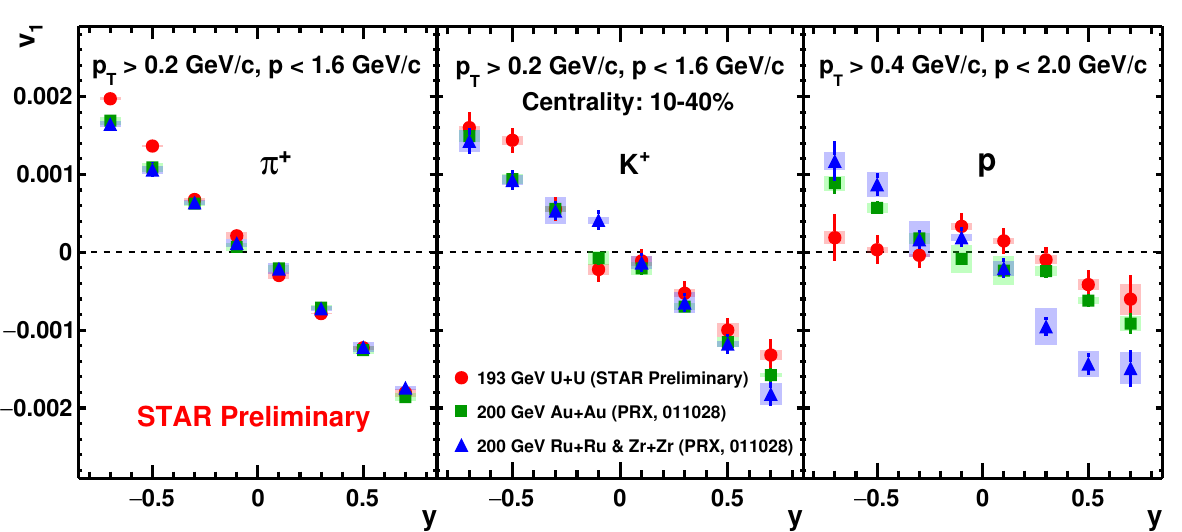}
\includegraphics[width=9.5 cm,clip]{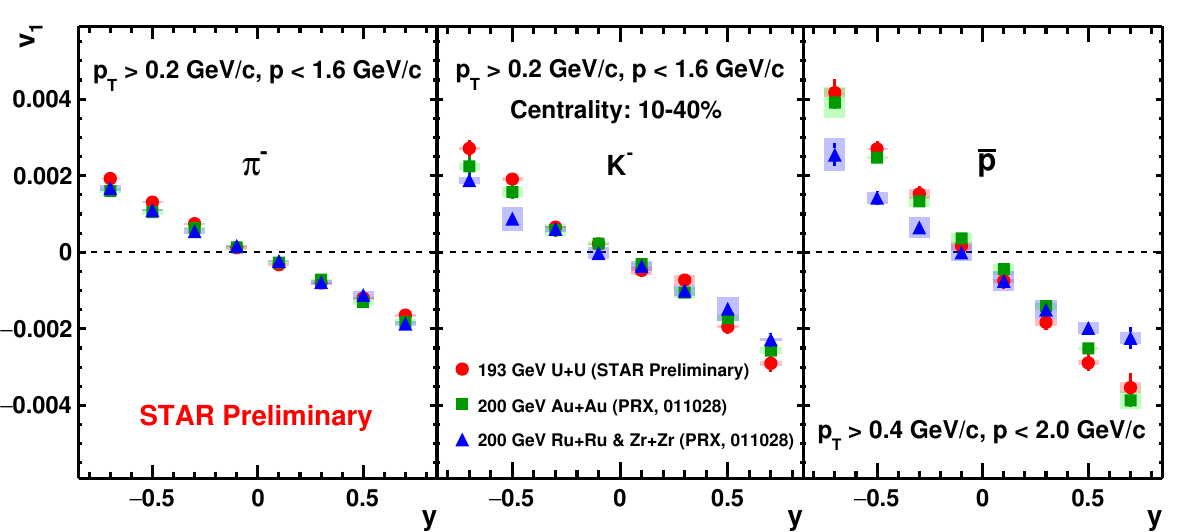}
\vspace{-5pt} 
\caption{$v_1$ as a function of rapidity for $\pi^\pm$, $K^\pm$, protons and anti-protons in U+U, Au+Au and isobar (Ru+Ru and Zr+Zr) collisions at $\sqrt{s_{NN}}$ = 193 and 200 GeV \cite{4}. Transverse momentum ($p_{T}$) and total momentum in collision center-of-mass frame ($p$) are specified in figure legends.}
\label{fig-1}       
\end{figure}
\vspace{-10pt} 
Figure 1 shows STAR measurement of directed flow as a function of rapidity in mid-central (10-40\%) U+U collisions for $\pi^\pm$, $K^\pm$ and $p(\bar{p}$). The slope ($dv_1/dy$) is extracted using a linear fit within rapidity (-0.8 < $y$ < 0.8). Moreover, $v_1(y)$ measured in this work for U+U is also compared with published results in Au+Au and isobar collisions at 200 GeV \cite{4}.
\clearpage
\begin{figure}[h]
\centering
\includegraphics[width=9.8cm,clip]{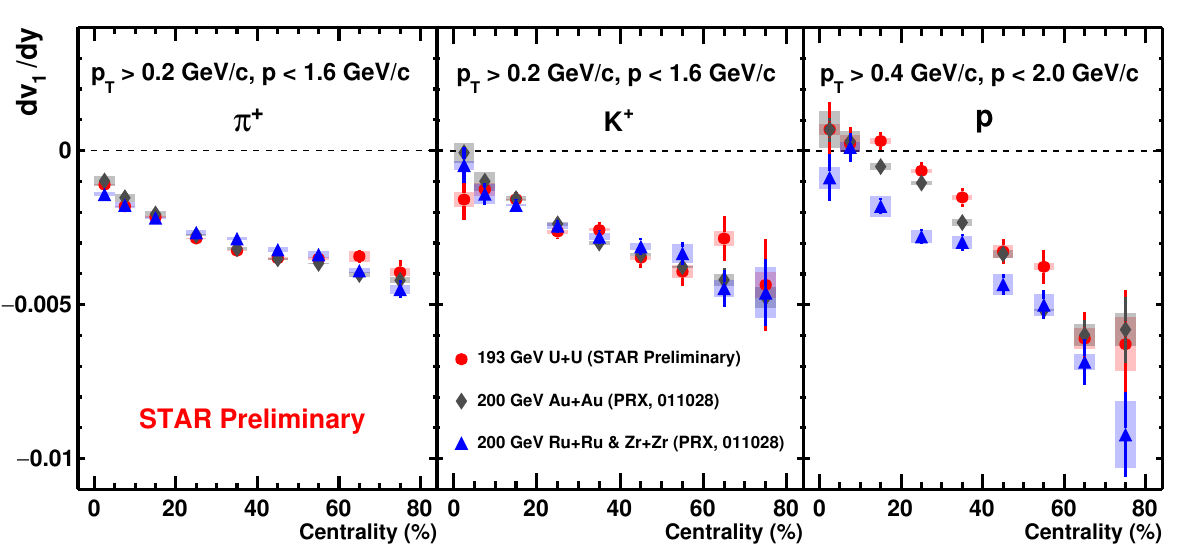}
\includegraphics[width=9.8cm,clip]{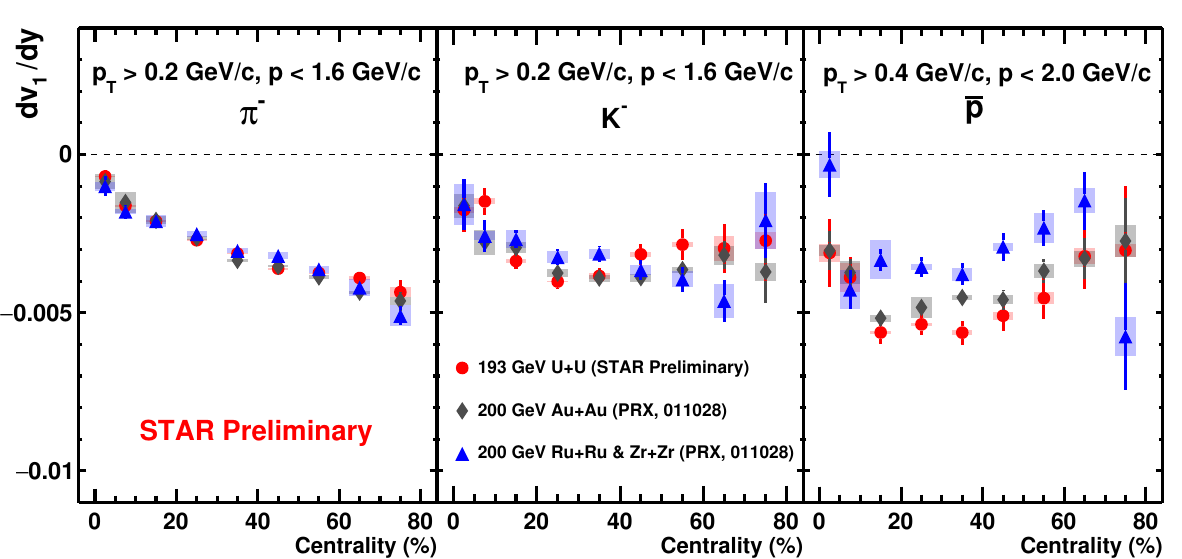}
\vspace{-5pt} 
\caption{$dv_1/dy$ as a function of centrality for $\pi^\pm$, $K^\pm$, protons and anti-protons in U+U, Au+Au and isobar (Ru+Ru and Zr+Zr) collisions at $\sqrt{s_{NN}}$ = 193 and 200 GeV \cite{4}.}
\includegraphics[width=9.8cm,clip]{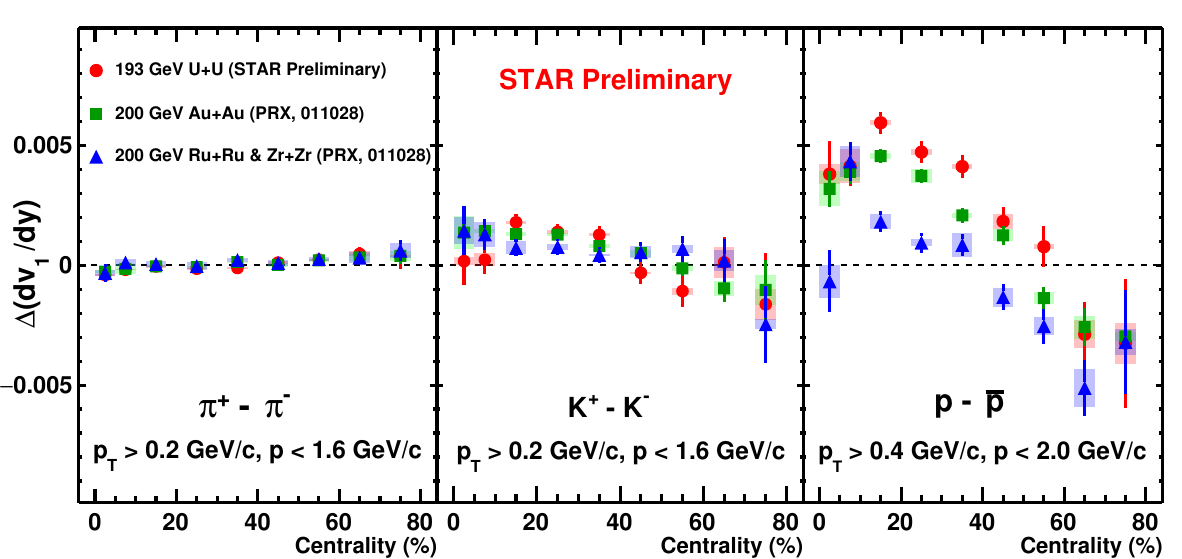}
\vspace{-5pt} 
\caption{$\Delta(dv_1/dy)$ as a function of centrality for pions, kaons and protons in U+U, Au+Au and isobar (Ru+Ru and Zr+Zr) collisions at $\sqrt{s_{NN}}$ = 193 and 200 GeV \cite{4}.}
\label{fig-1}       
\end{figure}
\vspace{-10pt} 
Figure 2 presents $(dv_1/dy)$ slope as a function of centrality for $\pi^\pm$, $K^\pm$ and $p(\bar{p}$). A significant splitting in $(dv_1/dy)$ slope as well as $dv_1/dy$ dependence on system size is observed for proton and antiproton in the mid-central (10-40\%) collisions among the three different collision systems. In contrary, no significant $dv_1/dy$ splitting is observed for pions and kaons and the data points are found consistent between different colliding species.
\medbreak
Figure 3 shows the difference of $(dv_1/dy)$ slope between positively and negatively charged particles as a function of centrality. Proton $\Delta$$(dv_1/dy)$ shows a clear ordering in system size in the mid-central (10-40\%) collisions as well as a sign change in peripheral (50-80\%) collisions. This sign change for the proton $\Delta$$(dv_1/dy)$ is consistent with naive expectations of transported quarks + electromagnetic effects. No significant $(dv_1/dy)$ splitting between $\pi^{+}$ and $\pi^{-}$ is observed in U+U collisions at 193 GeV similar to Au+Au and isobar collisions at 200 GeV. A similar but less obvious trend, compared to pion $\Delta$$(dv_1/dy)$, is found for kaon $\Delta$$(dv_1/dy)$ and the data points are consistent within uncertainties.
\vspace{-5pt} 
\medbreak
\begin{figure}[h]
\centering
\includegraphics[width=10cm,clip]{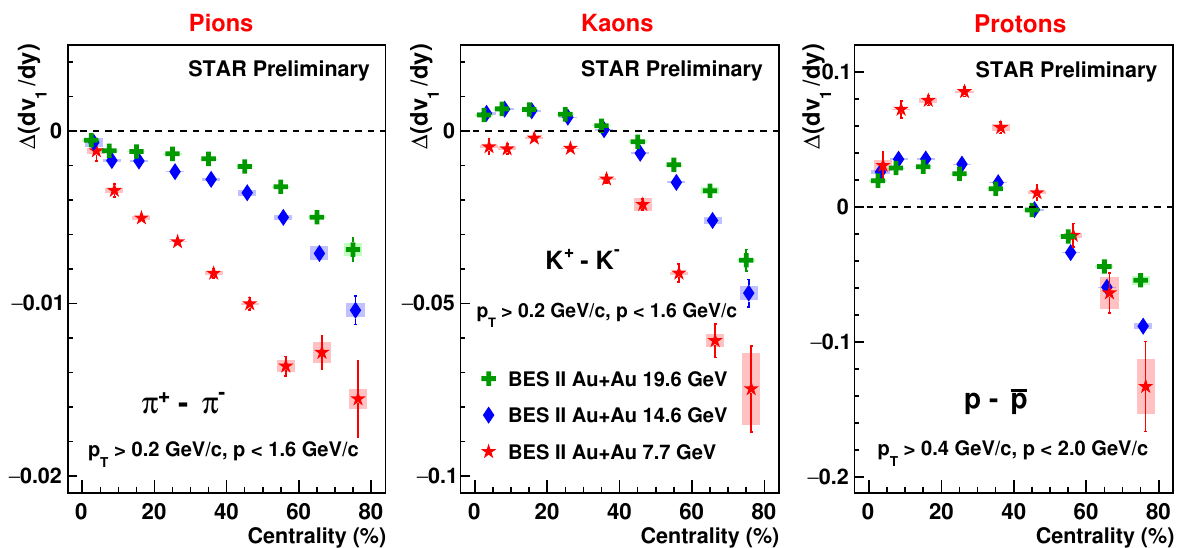}
\vspace{-5pt} 
\caption{$\Delta(dv_1/dy)$ as a function of centrality for pions, kaons and protons in Au+Au collisions at $\sqrt{s_{NN}}$ = 19.6, 14.6 and 7.7 GeV.}
\label{fig-1}       
\end{figure}
\vspace{-10pt} 
Figure 4 presents the beam energy dependence of $\Delta(dv_1/dy)$ as a function of centrality for pions, kaons and protons in Au+Au collisions at $\sqrt{s_{NN}}$ = 7.7 – 19.6 GeV. The $\Delta(dv_1/dy)$ for protons become more negative in the peripheral collisions as the beam energy is reduced, consistent with the expectation from the dominance of (Faraday + Coulomb) effect \cite{8}.


\section{Summary}
We have presented a measurement of $v_1$ and $\Delta(dv_1/dy)$ in U+U collisions at 193 GeV as well as in Au+Au collisions  at 7.7 – 19.6 GeV using STAR experiment. The positive $\Delta(dv_1/dy)$ for kaons and protons in central collisions can be attributed to the contributions from transported quarks, while the notable negative values of $\Delta(dv_1/dy)$, observed in the peripheral collisions, are consistent with the effects of the electromagnetic field, primarily due to Faraday induction combined with the Coulomb effect \cite{1, 2}. The $v_1$ dependence on system size can be due to the presence of extremely strong magnetic fields at very early stages of heavy-ion collisions, however it needs more exploration and theoretical input. Additionally, the stronger charge splitting observed in Au+Au collisions at lower BES-II energies \cite{8} support the idea that the electromagnetic field decays more slowly at lower collision energies. Other mechanisms such as effect from baryon transport and its inhomogeneities are under investigation \cite{10}. 

%
%
%


\end{document}